\documentclass[twocolumn,prc,aps,floatfix]{revtex4}
\usepackage[dvips]{graphicx}
\begin{document}
\title{Comment on ``Black hole constraints on varying fundamental constants''}
\author{V.V. Flambaum}
\affiliation{
 School of Physics, The University of New South Wales, Sydney NSW
2052, Australia
}
\date{\today}
\begin{abstract}
In the Letter \cite{Gibbon} (also \cite{DDL}) there is a  claim that the
generalised second law of thermodynamics (entropy increase)
 for black holes provides
some limits on the rate of  variation of the fundamental constants
of nature (electric charge $e$, speed of light $c$, etc.). 
We have come to a different conclusion.
 The results in  \cite{Gibbon,DDL} are based on
 assumption that mass of a black hole does not change without
 radiation and accreation.
 We present arguments showing that  this assumption is incorrect
and give an estimate of  the black hole mass variation due to
 $\alpha=e^2/\hbar c$ variation using entropy
(and quantum energy level) conservation in an adiabatic process.
No model-independent limits on the variation of the fundamental constants
are derived from the second law of thermodynamics.
\end{abstract}
\maketitle

It is convenient to present the dimensionless entropy
 of a charged black hole  \cite{entropy}
 in terms of dimensionless parameters:
\begin{equation}
\label{S}
S=\pi [\mu + \sqrt{\mu^2 - Z^2 \alpha}]^2 
\end{equation}
Here $\mu=M/M_P$, $M$ is the black hole mass,$M_P=(\hbar c/G)^{1/2}$ is the
Plank mass, $Ze$ is the black hole charge,
the Boltzmann constant $k=1$
(for a rotating black hole\\ 
$S=\pi ([\mu + \sqrt{\mu^2 - Z^2 \alpha - J(J+1)/\mu^2}]^2 +J(J+1)/\mu^2)$).
This expression does not contain explicitly any parameters which have 
dimension
% \cite{note}
 (speed of light, proton electric charge, etc.). 
Therefore, one may only discuss variation of two 
dimensionless parameters: mass of the black hole in units of the 
Plank mass ($\mu$) and $\alpha$.
% \cite{duff,rotation}. 
%Similar arguments were presented in  comments \cite{duff}.
%However, the question remains: can we extract any limitations on variation
%of $\mu(t)$ and $\alpha(t)$?
 In all known adiabatic processes the entropy $S$ is conserved.
It is natural to assume that this
is also valid for a  slow variation of the fundamental constants.
 Then eq. (\ref{S}) gives $\mu(t)$  in terms of  $\alpha(t)$ and constant $S$:
\begin{equation}
\label{mualphaS}
\mu=\frac{M}{M_P}=\frac{(S/\pi) +Z^2\alpha}{2 \sqrt{(S/\pi)}}
\end{equation}
 The event horizon area $A$ of the
 black hole is quantized \cite{bekenstein}. Because of the relation between
the entropy $S$ and the horizon area $A$ we obtain
the entropy quantization $S=(c^3/4 G\hbar) A = \pi \gamma \cdot n$,
where $\gamma$ is a numerical constant, $n$ is an integer.
%(in fact, in Ref. \cite{bekenstein} a neutral black hole 
%has been discussed; for the validity of this equation we possibly need 
%an extra condition $n >>1$).
 This gives us $\mu$ as a function of $\alpha$: 
%and the number of quanta $n$:
\begin{equation}
\label{mualpha}
\mu=\frac{M}{M_P}=\frac{\gamma \cdot n +Z^2\alpha}{2 \sqrt{\gamma \cdot n}}
\end{equation}
One may compare this result with that for the hydrogen atom
where we have the electron  energy levels
 $E_n/m_e c^2 \approx 1-\alpha^2/2n^2$.
An adiabatic variation of parameters do not cause any transitions between
non-equal levels, therefore the variation of atomic and black hole masses
is given by the stationary formulas  with $\alpha$ depending on time.
  Note that the variations of masses  do not contradict to the 
 energy conservation law since  atoms and  black holes are
 not  closed systems,
 they are  interacting with the Universe. Indeed, in theoretical models
the variation is driven by an  evolution of some scalar field, and  energy
of this field must be taken into account.
The entropy in the adiabatic  case does not change.
 Therefore, the time dependence
of $\alpha$ does not lead to any specific problems for the 
black holes, it just gives us the dependence $\mu(t)$.

One could suggest (see e.g. \cite{DDL}) that the variation 
$\alpha (t)$ may lead to a  negative expression
under the square root sign in eq. (\ref{S}) (which also appears
in the formulas for the horizon area and temperature), and this may give
a limit on the allowed variation. Here again an atomic
analogy may be useful.
 The energy of the $1s$ level in the Dirac
equation  is $E= m c^2 \sqrt{1- Z^2 \alpha^2}$.
If we  increase $\alpha$ beyond $1/Z$
the expression under the square root becomes negative and the stationary state
$1s$ disappears.
% (for the nucleus of any finite size it does not happen at
%energy $E=0$ but the problem appears again for $E= -mc^2$).
 This only means that the system becomes non-stationary.
 The strong Coulomb field
creates an electron-positron pair, the positron goes to infinity
and the nuclear charge reduces to a sub-critical value ($Z$ to $Z-1$).
Therefore, the possibility of a negative expression 
under the square root sign in the electron energy level
 or eq. (\ref{S}) for the black hole entropy
 does not mean that certain variations of $\alpha$ are forbidden.

  An inclusion of the Hawking radiation and accreation leads to an
increase of entropy, therefore, it does not  violate the second law
of thermodynamics, at least for the adiabatic variation of the fundamental
constants. For a non-adiabatic variation one should accurately take
into account contributions of scalar fields driving the variation.

\end{document}